\documentstyle[12pt,epsfig]{article}

\begin{document}
\begin {center}
{\bf {\Large
Color transparency of $K^+$ meson produced in the inclusive
$(e,e^\prime)$ reaction on nuclei
} }
\end {center}
\begin {center}
Swapan Das \footnote {email: swapand@barc.gov.in} \\
{\it Nuclear Physics Division,
Bhabha Atomic Research Centre,  \\
Trombay, Mumbai: 400085, India \\
Homi Bhabha National Institute, Anushakti Nagar,
Mumbai: 400094, India }
\end {center}

\begin {abstract}
Color transparency is studied for the $K^+$ meson produced due to large
four-momentum transfer in the $(e,e^\prime)$ reaction on nuclei. The
variation of the $K^+$ meson color transparency ($K^+$CT) with the photon
virtuality, and kaon momentum is investigated. The calculated results for
$K^+$CT are compared with the data reported from Jefferson Laboratory.
\end {abstract}

Keywords:
color transparency, $K^+$ meson production, electronuclear reaction


\section {Introduction}

The characteristics of the $K^+$ meson are distinctly different from those
of other hadrons. The interaction of this meson with a nucleon is
relatively weak, free from resonance structure, and that varies smoothly
with the energy. Because of these behaviors, the $K^+$ meson can be thought
as a pertinent probe to investigate specific properties of a nucleus.
The
scattering of this meson from a nucleus can provide informations
complementary to those obtained from the electron-nucleus scattering, since
both are weakly interacting probes unlike to the conventional strongly
interacting hadronic probe. Additionally, the $K^+$ meson can open the
avenue for studying the strangeness degrees of freedom in the nuclear
reaction.

Despite the $K^+$ meson possesses such useful properties, the description of
the $K^+$ meson-nucleus scattering from the elementary $K^+$ meson-nucleon
scattering is not successful. The calculated $K^+$ meson-nucleus cross
section \cite{Siegel, Das} shows discrepancy with the data \cite{Bugg}.
Even
the calculated nuclear transparency \cite{Siegel, Das}, defined by the ratio
of the total cross sections for a nucleus and deuteron (where the
uncertainties are expected to cancel), underestimates the experimental
results \cite{Bugg}.
Because
of the failure of the conventional nuclear physics to explain the quoted
data, several exotic mechanisms for the $K^+$ meson-nucleus scattering
have been proposed.
These
include modification of the in-medium nucleon size/mass \cite{Siegel, Brown},
medium modification of the elementary $K^+$ meson-nucleon scattering
amplitude \cite{Chauhan}, virtual pion contribution \cite{Akul}, mesons
exchange currents \cite{Jiang}, long range correlations \cite{Garcia}, and
various other mechanisms \cite{Caillon}.

The electroproduction of the $K^+$ meson from a nucleus provides an
alternate tool to investigate the $K^+$ meson-nucleus scattering, and also
to explore the propagation of this meson through the nucleus.
The
dependence of the $K^+$ meson transparency on the nuclear mass number $A$
and four-momentum transfer squared $Q^2$ (i.e., photon virtuality) in the
$A(e,e^\prime)$ reaction has been reported from Jefferson Laboratory
\cite{Nuru}.
The
data available from this Laboratory renew the interest in looking for
whether the discrepancy between the theoretical and experimental results
also exist for the $K^+$ meson production reaction.

It has been shown later that the conventional nuclear physics calculation
fails to reproduce the electroproduced $K^+$ meson nuclear transparency
data \cite{Nuru}, similar to that occurred for the $K^+$ meson-nucleus
scattering data.
Therefore,
color transparency \cite{Frank} of the $K^+$ meson is envisaged as high
four-momentum transfer is involved to produce this meson. Color transparency
of a hadron describes the enhancement in its transparency in the nucleus.
The
quoted enhancement originates because of the reduction of the hadron-nucleon
interaction (or total cross section $\sigma^{hN}_t$) while the hadron
undergoes large four-momentum transfer during its passage through the nucleus
\cite{Howell}.

The high momentum transfer $Q^2$ associated with a hadron reduces its
transverse size (i.e., $d_\perp \sim 1/\sqrt{Q^2}$). The reduced (in size)
hadron is referred as small-size or pointlike configuration (PLC).
According
to Quantum Chromodynamics, a color singlet PLC has reduced interaction with
nucleons in the nucleus because the sum of the gluon emission amplitudes
cancel \cite{Dutta2}.
The
interaction of PLC with the nucleon increases, as the PLC expands to the
size of the physical hadron during its passage upto the length called hadron
formation length $l_h$ \cite{Farrar}:
\begin{equation}
l_h= \frac{2k_h}{\Delta M^2},
\label{Lh}
\end{equation}
where $k_h$ is the momentum of the hadron in the laboratory frame.
$\Delta M^2$ is related to the mass difference between the hadronic states
originating due to the (anti)quarks fluctuation in PLC. The value of
$\Delta M^2$ is very much uncertain, ranging from 0.25 to 1.4 GeV$^2$
\cite{Farrar}.
Therefore,
the effective hadron-nucleon total cross section $\sigma^{hN}_{t,eff}$ in
the nucleus depends on both $Q^2$ and $l_h$, i.e.,
$\sigma^{hN}_t \to \sigma^{hN}_{t,eff} (Q^2, l_h)$ in the nucleus.

First experiment to search color transparency of the proton done at
Brookhaven National Laboratory \cite{Carr} in the high momentum transfer
$(p,pp)$ reactions on nuclei yielded null result, and that was corroborated
in the later experiments \cite{Mardor2}.
According
to Landshoff mechanism \cite{Land}, three well-separated quarks $(qqq)$ in
one proton interact with those in another proton exchanging three gluons
\cite{Rals}.
Color
transparency is not also seen in the $A(e,e^\prime p)$ experiment done
at Stanford Linear Accelerator Center \cite{Neil} and Jefferson Laboratory
\cite{Garr}.
Therefore, it appears that the PLC required for color transparency is
unlikely to form for the $qqq$ system, e.g., proton.

Color transparency is unambiguously reported from Fermi National
Accelerator Laboratory \cite{aitala} in the experiment of the nuclear
diffractive dissociation of pion (of 500 GeV/$c$) to dijets.
Since
meson is a bound state of two quarks (i.e., quark-antiquark $q{\bar q}$),
PLC of it is more probable than that of baryon, a $qqq$ object.
Color
transparency of the meson $M$ is also found in the photon \cite{Dutta} and
electron induced nuclear experiments \cite{Clasie, Aira}. In the latter
reaction, i.e., $A(e,e^\prime M)$ reaction,
the
length of the $q{\bar q}$ (of mass $m_{q{\bar q}}$) fluctuation to a meson
in the virtual photon $\gamma^*$ (of energy $\nu$ and virtuality $Q^2$) is
described by the coherence length (CL) \cite{Aira}:
$l_c= \frac{2\nu}{Q^2+m^2_{q{\bar q}}}$.
$T_A$
varies with $l_c$ in absence of color transparency. Therefore, $l_c$ must be
kept fixed to observe color transparency \cite{Adams}. Several authors have
studied the $\rho$ meson color transparency ($\rho$CT), and the effect of CL
on $\rho$CT in the energy region available at Jefferson Laboratory (JLab)
\cite{Howell, Kopel}.

The nuclear transparencies for both $\pi^+$ \cite{Clasie} and $K^+$
\cite{Nuru} mesons versus the photon virtuality $Q^2$ (of few GeV) in the
$A(e,e^\prime)$ reaction were measured at JLab. The data for both mesons
could not be understood by using the conventional nuclear physics.
The
pionic color transparency ($\pi$CT) in above reaction is shown to occur by
Kaskulov et al., \cite{Kasku} in their calculation done using couple channel
BUU transport model.
Larson et al.,
\cite{Larson} illustrates the dependence of $\pi$CT on the pion momentum
and momentum transfer to nucleus (instead of $Q^2$) using semiclassical
formula for the nuclear transparency. Cosyn et al., \cite{cosyn} calculated
CT and short-range correlation in the pion photo- and electro-production
from nuclei.
The
calculated results due to Larinov et al., \cite{Lari} show $\pi$CT in the
$A(\pi^-,l^+l^-)$ process at $\approx 20$ GeV/$c$, which can be measured at
the forth-coming facilities in Japan Proton Accelerator Research Complex
(cf. \cite{Kumano} and the references therein). This study provides the
information analogous to that obtained in the $A(e,e^\prime \pi)$ reaction.
Miller
and Strikman \cite{Miller} show large color transparency in the pionic
knockout of proton off nuclei, i.e., $A(\pi,\pi p)$ reaction, at the energy
500 GeV available at CERN COMPASS experiment. The recent development of color
transparency and its future direction are discussed in Ref.~\cite{Dutta2}.
As
illustrated later, the $K^+$ meson nuclear transparency calculated
considering color transparency of this meson reproduce well the data reported
from Jlab \cite{Nuru}.

\section {Formalism}

The nuclear transparency $T_A (K^+)$ of the $K^+$ meson, including the small
angle kaon-nucleus elastic scattering, can be written as
\begin{equation}
T_A (K^+)= \frac{ \sigma_{in}^{K^+A} + \Delta \sigma_{el}^{K^+A} }
{A\sigma_t^{K^+N}},
\label{TApF}
\end{equation}
where $\sigma_{in}^{K^+A}$ is the inelastic $K^+$ meson-nucleus cross section
and $\Delta \sigma_{el}^{K^+A}$ represents the small angle kaon-nucleus
elastic scattering cross section. $\sigma_t^{K^+N}$ is the elementary $K^+$
meson-nucleon total cross section:
$\sigma_t^{K^+N}$ =
$\frac{Z}{A} \sigma_t^{K^+p}$ + $\frac{A-Z}{A} \sigma_t^{K^+n}$.
In
the considered kaon momentum region, i.e., $\approx 3-10$ GeV/$c$, the energy
dependent experimentally determined values of $\sigma_t^{K^+p}$ and
$\sigma_t^{K^+n}$, as given in Ref.~\cite{Bail}, are almost equal, i.e.,
$\sigma_t^{K^+p} \simeq \sigma_t^{K^+n}$.

The inelastic cross section $\sigma_{in}^{K^+A}$ in Eq.~(\ref{TApF}),
according to the optical approach in Glauber model \cite{Glaub}, is given
by
\begin{equation}
\sigma_{in}^{K^+A} =
\int d{\bf b} \left [ 1- \left |e^{i\chi_{OK} ({\bf b})} \right |^2 \right ].
\label{Inxt}
\end{equation}
$\chi_{OK} ({\bf b})$ denotes the optical phase-shift for the $K^+$ meson,
i.e.,
\begin{equation}
\chi_{OK} ({\bf b})
= -\frac{1}{v} \int^{+\infty}_{-\infty} dz V_{OK}({\bf b}, z),
\label{PhKm}
\end{equation}
where $V_{OK}({\bf r})$ is the $K^+$ meson-nucleus optical potential
\cite{Glaub}:
\begin{equation}
-\frac{1}{v} V_{OK}({\bf r})=
\frac{1}{2} (\alpha_{K^+N}+i) \sigma_t^{K^+N} \varrho({\bf r}).
\label{VOKm}
\end{equation}
$\alpha_{K^+N}$ represents the ratio of the real to imaginary part of the
$K^+$ meson-nucleon elastic scattering amplitude \cite{Bail}.
$\varrho({\bf r})$ describes the density distribution of the nucleus, i.e.,
it is to be normalized to the mass number of the nucleus.

Using Eqs.~(\ref{PhKm}) and (\ref{VOKm}), $\sigma_{in}^{K^+A}$ in
Eq.~(\ref{Inxt}) can be written as
\begin{eqnarray}
\sigma_{in}^{K^+A}
&=& \int d{\bf b} \int^{+\infty}_{-\infty} dz
    \left \{ -\frac{2}{v} Im V_{OK} ({\bf b}, z) \right \}
     exp \left [ \int^{+\infty}_z
     \left \{ \frac{2}{v} Im V_{OK} ({\bf b}, z^\prime) dz^\prime \right \}
     \right ]
    \nonumber  \\
&=& \sigma_t^{K^+N} \int d{\bf b} \int^{+\infty}_{-\infty} dz
    \varrho ({\bf b}, z)
     exp \left [ -\sigma_t^{K^+N} \int^{+\infty}_z
    \varrho ({\bf b},z^\prime) dz^\prime \right ].
\label{Inxt2}
\end{eqnarray}
The nuclear transparency defined by
$T_A$ = $\frac{\sigma_{in}^{K^+A}}{A\sigma_t^{K^+N}}$ refers to the
expression due to semiclassical model as mentioned in Ref.~\cite{Farrar},
i.e.,
\begin{equation}
T_A (K^+)
= \frac{1}{A} \int d{\bf r} \varrho ({\bf b}, z)
   exp [ -\sigma_t^{K^+N} \int^{+\infty}_z \varrho_A ({\bf b}, z^\prime)
   dz^\prime ].
\label{TApF2}
\end{equation}

The cross section $\Delta \sigma_{el}^{K^+A}$ in Eq.~(\ref{TApF}) is given
by
\begin{equation}
\Delta \sigma_{el}^{K^+A}
= \frac{ d\sigma^{K^+A}_{el} }{ d\Omega } \Delta \Omega
= |F_{K^+A}|^2 \Delta \Omega,
\label{SaeX}
\end{equation}
where $\Delta \Omega$ (=6.7 msr \cite{Clasie}) is the angular aperture of
the detector. $F_{K^+A}$ represents the kaon-nucleus elastic scattering
amplitude \cite{Glaub}:
\begin{eqnarray}
F_{K^+A}
&=& \frac{k_{K^+}}{2\pi i} \int d{\bf b} e^{i{\bf q_\perp.b}}
    \left [ e^{i\chi_{OK} ({\bf b})} -1 \right ],
\label{Scam}
\end{eqnarray}
where $q_\perp (= -k_{K^+} sin\theta_{K^+})$ represents the transverse
momentum transfer.

\section {Result and Discussions}

The nuclear transparency $T_A (K^+)$ of the $K^+$ meson produced in the
$(e,e^\prime)$ reaction on $d$, $^{12}$C, $^{63}$Cu and $^{197}$Au nuclei
has been calculated using Glauber model, where the experimentally determined
free space kaon-nucleon total cross section $\sigma^{K^+N}_t$ \cite{Bail}
has been used.
The
density distribution $\varrho (r)$ of deuteron $(d)$ nucleus is evaluated
using its wave function generated due to paris potential \cite{Lacombe}.
$\varrho$
for other nuclei, as extracted from the electron scattering data, is
tabulated in Ref.~\cite{Jager}. According to it, $\varrho$ for $^{12}$C
nucleus is described by the harmonic oscillator Gaussian form, whereas that
for other nuclei (i.e., $^{63}$Cu and $^{197}$Au) is illustrated by
two-parameter Fermi distribution.

The calculated results for $T_A (K^+)$ in Eq.~(\ref{TApF}) show that the
contribution of the small angle elastic scattering cross section, i.e.,
$\Delta \sigma_{el}^{K^+A}$, to $T_A (K^+)$ is negligible compared to that
of $\sigma^{K^+A}_{in}$.
It
should be mentioned that the kinematics of the experiment is so chosen that
the elastic scattering of kaon from the nucleus would be suppressed
\cite{Dutta2, Clasie}. Therefore, $\Delta \sigma_{el}^{K^+A}$ is neglected
to evaluate $T_A (K^+)$.

The dependence of the calculated $T_A (K^+)$ in Eq.~(\ref{TApF2}) on the
photon virtuality $Q^2$ are presented in Fig.~\ref{TQpK} along with the data
\cite{Nuru}. The short-dashed curves distinctly show that $T_A (K^+)$,
evaluated using $\sigma^{K^+N}_t$ in Glauber model, significantly
underestimate the electroproduction data for the $K^+$ meson nuclear
transparency.
Since
few GeV four-momentum transfer $Q^2$ is involved for the $K^+$ meson
production, color transparency of this meson is considered. According to it,
the effective kaon-nucleon total cross section $\sigma^{K^+N}_{t,eff}$ in the
nucleus is less than $\sigma^{K^+N}_t$ because of PLC formation of the
$K^+$ meson in the nucleus. Using quantum diffusion model \cite{Farrar},
$\sigma^{K^+N}_{t,eff}$ can be written as
\begin{equation}
\sigma^{K^+N}_{t,eff} (Q^2, l_h; l_z) =\sigma^{K^+N}_t
\left [
\left \{
\frac{l_z}{l_h} +\frac{n_q^2<k^2_t>}{Q^2} \left ( 1 -\frac{l_z}{l_h} \right )
\right \} \theta(l_h-l_z)
+ \theta(l_z-l_h) \right ],
\label{XCT}
\end{equation}
where  $n_q(=2)$ denotes the number of the valence quark-antiquark
$q{\bar q}$ in the meson. $k_t$ illustrates the transverse momentum of the
(anti)quark: $<k^2>^{1/2} =0.35$ GeV/c. $l_z$ is the path-length traversed
by the meson $(q{\bar q})$ after its production.
The
hadron formation length $l_h (\propto \frac{1}{\Delta M^2})$ is already
defined in Eq.~(\ref{Lh}). For $l_h=0$, i.e., $\Delta M^2=\infty$, the above
equation is reduced to $\sigma^{K^+N}_t$.

$T_A(K^+)$ evaluated using $\sigma^{K^+N}_{t,eff}$ in Eq.~(\ref{XCT}) is also
shown in Fig.~\ref{TQpK}. The dot-dot-dashed curves represent the calculated
$T_A(K^+)$ for $\Delta M^2 =1.4$ GeV$^2$, whereas the dot-dashed and solid
curves arise due to $\Delta M^2$ taken equal to 0.7 GeV$^2$ and 0.3 GeV$^2$
respectively.
The
calculated results illustrate $T_A(K^+)$ increases with $Q^2$ because of
$\sigma^{K^+N}_{t,eff}$. It is noticeable in the figure $T_A(K^+)$ evaluated
for $\Delta M^2 =0.3$ GeV$^2$ (solid curves) agrees well with the data
\cite{Nuru} for all nuclei. The calculated results describe color
transparency of the $K^+$ meson produced in the $(e,e^\prime)$ reaction on
nuclei.

The calculated kaonic transparency ratio $T_{A/d}(K^+)$ with respect to
deuteron, i.e., $T_{A/d}(K^+) =T_A(K^+)/T_d(K^+)$, is compared with the data
\cite{Nuru} in Fig.~\ref{TQdK}. This ratio includes the contribution of the
$K^+$ meson production from neutron, and Fermi-motion correction.
The
curves appearing in this figure represent those illustrated in
Fig.~\ref{TQpK}. Color transparency of $K^+$ meson is also distinctly
visible in Fig.~\ref{TQdK}, as the calculated $T_{A/d}(K^+)$ for
$\Delta M^2 =0.3$ GeV$^2$ (solid curves) reproduce the data \cite{Nuru}
reasonably well.

The kaon momentum $k_{K^+}$ dependent $T_A(K^+)$ is shown in Fig.~\ref{TKdK}
for different values of $\Delta M^2$. In fact, it describes the variation of
$T_A(K^+)$ with the hadron formation length, mentioned in Eq.~(\ref{Lh}),
of the $K^+$ meson.
The
range of $k_{K^+}$ is considered up to 9.5 GeV/$c$ which would be accessible
in the kinematics of 11 GeV Jlab experiment. The calculated results show
$T_A(K^+)$ increases with the kaon momentum.

\section {Conclusion}

The nuclear transparency $T_A$ and the transparency ratio, i.e.,
$T_{A/d}= T_A/T_d$, of the $K^+$ meson produced in the $A(e,e^\prime)$
reaction have been calculated to look for color transparency of the $K^+$
meson.
The
calculated results show that both $T_A$ and $T_{A/d}$ evaluated using the
free space $K^+$ meson-nucleon cross section $\sigma_t^{K^+N}$ in Glauber
model significantly underestimate the data reported from JLab.
The
inclusion of the $K^+$ meson color transparency in Glauber model (i.e., the
use of kaon-nucleon effective cross section $\sigma^{K^+N}_{t,eff}$, instead
of $\sigma_t^{K^+N}$, arising due to the formation of the pointlike
configuration of kaon) leads to the enhancement in the calculated
transparency.
The
calculated $T_A$ and $T_{A/d}$ due to $\sigma^{K^+N}_{t,eff}$ (evaluated for
$\Delta M^2 =0.3$ GeV$^2$) are well accord with the data, i.e., those
describe color transparency of the $K^+$ meson produced in the
electronuclear reaction.

\section {Acknowledgement}

The author thanks D. Dutta for sending the kinematic of the reaction, and
appreciates the anonymous referee for the comments which improve the quality
of the paper.

\newpage

\begin{figure}[h]
\begin{center}
\centerline {\vbox {
\psfig{figure=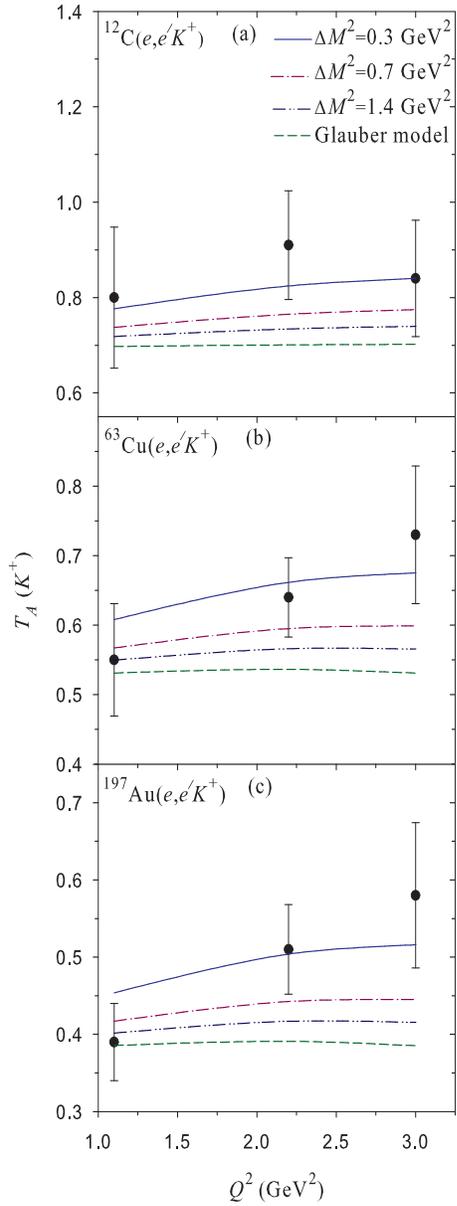,height=16.0 cm,width=06.0 cm}
}}
\caption{
The photon virtuality $Q^2$ dependent nuclear transparency $T_A(K^+)$ of
the $K^+$ meson produced in the $(e, e^\prime)$ reaction on the nucleus.
The dashed curves denote $T_A(K^+)$ calculated using Glauber model. Other
curves illustrate the $K^+$ meson color transparency for different values of
$\Delta M^2$, see in text. The data are taken from Ref.~\cite{Nuru}.
}               
\label{TQpK}
\end{center}
\end{figure}

\begin{figure}[h]
\begin{center}
\centerline {\vbox {
\psfig{figure=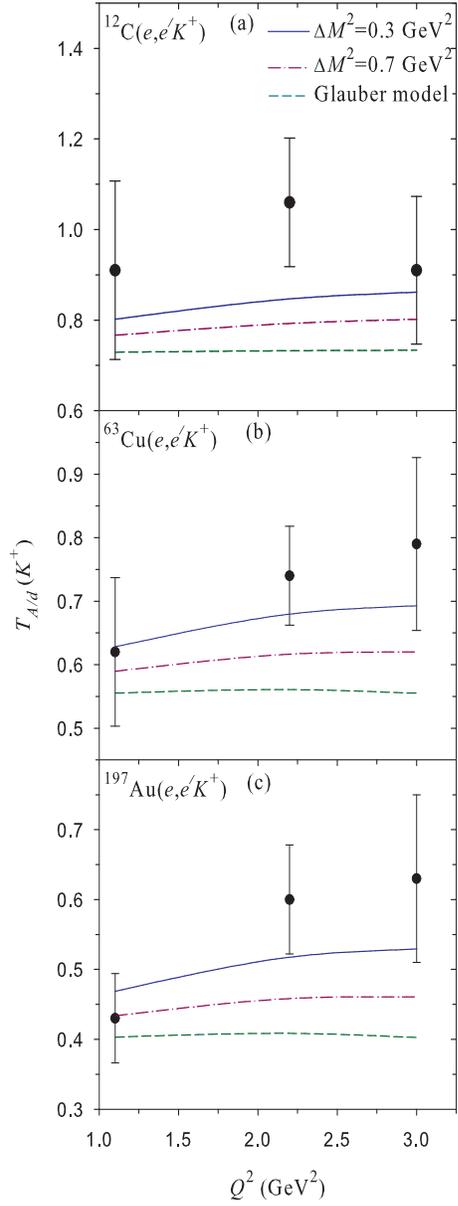,height=16.0 cm,width=06.0 cm}
}}
\caption{
The $K^+$ meson transparency ratio with respect to deuteron
$T_{A/d}= T_A/T_d$. The curves appearing in it represent those stated in
Fig.~\ref{TQpK}. The data are taken from Ref.~\cite{Nuru}.
}               
\label{TQdK}
\end{center}
\end{figure}

\begin{figure}[h]
\begin{center}
\centerline {\vbox {
\psfig{figure=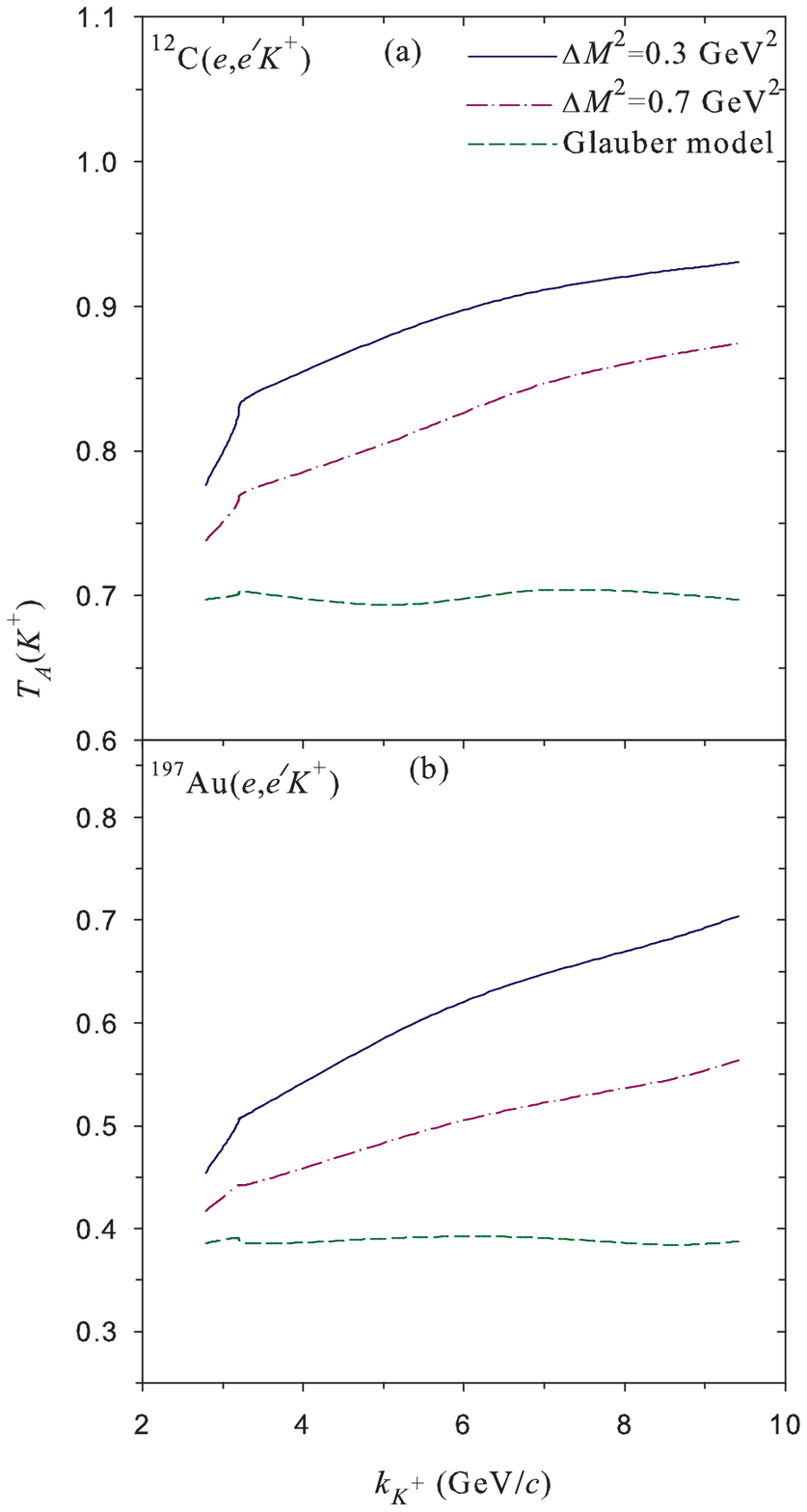,height=16.0 cm,width=08.0 cm}
}}
\caption{
The calculated results showing the variation of $T_A(K^+)$ with the $K^+$
meson momentum $k_{K^+}$. The curves appearing in it are described in
Fig.~\ref{TQpK}.
}               
\label{TKdK}
\end{center}
\end{figure}

{\bf Figure Captions}
\begin{enumerate}

\item
(color online).
The photon virtuality $Q^2$ dependent nuclear transparency $T_A(K^+)$ of
the $K^+$ meson produced in the $(e, e^\prime)$ reaction on the nucleus.
The dashed curves denote $T_A(K^+)$ calculated using Glauber model. Other
curves illustrate the $K^+$ meson color transparency for different values of
$\Delta M^2$, see in text. The data are taken from Ref.~\cite{Nuru}.

\item
(color online).
The $K^+$ meson transparency ratio with respect to deuteron
$T_{A/d}= T_A/T_d$. The curves appearing in it represent those stated in
Fig.~\ref{TQpK}. The data are taken from Ref.~\cite{Nuru}.

\item
(color online).
The calculated results showing the variation of $T_A(K^+)$ with the $K^+$
meson momentum $k_{K^+}$. The curves appearing in it are described in
Fig.~\ref{TQpK}.

\end{enumerate}

\newpage

\newpage

\newpage


\newpage
\begin{thebibliography}{100}

\bibitem{Siegel}
P. B. Siegel, W. B. Kaufmann and W. R. Gibbs,
Phys. Rev. C {\bf 31} (1985) 2184.

\bibitem{Das}
S. Das and A. K. Jain, Pramana {\bf 61} (2003) 1015.
X. Li, L. E. Write, and D. J. Bennhold, Phys. Rev. C {\bf 45} (1992) 2011.
M. J. P\a'ez and R. H. Landau, Phys. Rev. C {\bf 24} (1981) 1120.
P. B. Siegel, W. B. Kaufmann and W. R. Gibbs,
Phys. Rev. C {\bf 30} (1984) 1256.

\bibitem{Bugg}
D. Bugg et al., Phys. Rev. {\bf 168} (1968) 1466;
D. Marlow et al., Phys. Rev. C {\bf 25} (1982) 2619;
Y. Mardor et al., Phys. Rev. Lett. {\bf 65} (1990) 2110;
R. A. Krauss et al., Phys. Rev. C {\bf 46} (1992) 655;
R. Swafta et al., Phys. Lett. B {\bf 307} (1993) 293;
R. Weiss et al., Phys. Rev. C {\bf 49} (1994) 2569.

\bibitem{Brown}
G. E. Brown, C. B. Dover, P. B. Siegel and W. Weiss,
Phys. Rev. Lett. {\bf 60} (1988) 2723.

\bibitem{Chauhan}
D. Chauhan and Z. A. Khan, Phys. Rev. C {\bf 85} (2012) 067602.

\bibitem{Akul}
S. V. Akulinichev, Phys. Rev. Lett. {\bf 68} (1992) 290.

\bibitem{Jiang}
M. F. Jiang and D. S. Koltun, Phys. Rev. C {\bf 46} (1992) 2462.

\bibitem{Garcia}
C. Garc$\i'$a-Recio, J. Nieves and E. Oset,
Phys. Rev. C {\bf 51} (1995) 237.

\bibitem{Caillon}
J. C. Caillon and J. Labarsouque, Phys. Rev. C {\bf 45} (1992) 2503;
Phys. Lett. B {\bf 295} (1992) 21; J. Phys. G {\bf 19} (1993) L117;
Phys. Lett. B {\bf 311} (1993) 19.

\bibitem{Nuru}
Nuruzzaman et al., Phys. Rev. C {\bf 84} (2011) 015210.

\bibitem{Frank}
L. Frankfurt, G. A. Miller and M. Strikman,
Comments Nucl. Part. Phys. {\bf 21} (1992) 1; 
Annu. Rev. Nucl. Part. Sci. {\bf 45} (1994) 501;
L. Frankfurt, and M. Strikman, Phys. Rep. {\bf 160} (1988) 235;
P. Jain, B. Pire and J. P. Ralston, Phys. Rep. {\bf 271} (1996) 67.

\bibitem{Howell}
G. T. Howell and G. A. Miller, Phys. Rev. C {\bf 88} (2013) 035202.

\bibitem{Dutta2}
D. Dutta, K. Hafidi and M. Strikman,
Prog. Part. Nucl. Phys. {\bf 69} (2013) 1.

\bibitem{Farrar}
G. R. Farrar, H. Liu, L. L. Frankfurt and M. I. Strikman,
Phys. Rev. Lett. {\bf 61} (1988) 686.

\bibitem{Carr}
A. S. Carroll et al., Phys. Rev. Lett. {\bf 61} (1988) 1698.

\bibitem{Mardor2}
I. Mardor et al., Phys. Rev. Lett. {\bf 81} (1998) 5085;
A. Leksanov et al., Phys. Rev. Lett. {\bf 87} (2001) 212301.

\bibitem{Land}
P. V. Landshoff, Phys. Rev. D {\bf 10} (1974) 1024.

\bibitem{Rals}
J. P. Ralston and B. Pire, Phys. Rev. Lett. {\bf 61} (1988) 1823.

\bibitem{Neil}
T. G. O'Neill et al., Phys. Lett. B {\bf 351} (1995) 87;
N. C. R. Makins et al.,  Phys. Rev. Lett. {\bf 72} (1994) 1986.

\bibitem{Garr}
K. Garrow et al., Phys. Rev. C {\bf 66} (2002) 044613.

\bibitem{aitala}
E. M. Aitala et al., Phys. Rev. Lett. {\bf 86} (2001) 4773.

\bibitem{Dutta}
D. Dutta et al. (E94104 Collaboration), Phys. Rev. C {\bf 68} (2003)
021001R.

\bibitem{Clasie}
B. Clasie et al., Phys. Rev. Lett. {\bf 99} (2007) 242502;
X. Qian et al., Phys. Rev. C {\bf 81} (2010) 055209.

\bibitem{Aira}
A. Airapetian et al. (HERMES Collaboration),
Phys. Rev. Lett. {\bf 90} (2003) 052501;
L. El Fassi et al., Phys. Lett. B {\bf 712} (2012) 326.

\bibitem{Adams}
M. R. Adams et al., Phys. Rev. Lett. {\bf 74} (1995) 1525.

\bibitem{Kopel}
B. Z. Kopeliovich, J. Nemchik and I. Schmidt,
Phys. Rev. C {\bf 76} (2007) 015205;
L. Frankfurt, G. A. Miller and M. Strikeman,
Phys. Rev. C {\bf 78} (2008) 015208.

\bibitem{Kasku}
M. M. Kaskulov, K. Gallmeister and U. Mosel,
Phys. Rev. C {\bf 79} (2009) 015207.

\bibitem{Larson}
A. Larson, G. A. Miller and M. Strikman,
Phys. Rev. C {\bf 74} (2006) 018201.

\bibitem{cosyn}
W. Cosyn, M. C. Mart$\i'$nez and J. Ryckebusch,
Phys. Rev. C  {\bf 77} (2008) 034602.

\bibitem{Lari}
A. B. Larionov, M. Strikman and M. Bleicher,
Phys. Rev. C {\bf 93} (2016) 034618.

\bibitem{Kumano}
S. Kumano, in 21st International Symposium on Spin Physics (SPIN 2014)
Beijing, China, October 20-24, 2014: arXiv 1504.05264 [hep-ph].

\bibitem{Miller}
G. A. Miller and M. Strikman, Phys. Rev. C {\bf 82} (2010) 025205.

\bibitem{Bail}
P. Baillon et al., Nucl. Phys. B {\bf 105} (1976) 365;
ibid, {\bf 134} (1978) 31;
M. Tanabashi et al., (Particle Data Group),
Phys. Rev. D {\bf 98} (2018) 030001;
http://pdg.lbl.gov./xsect/contents.html.

\bibitem{Glaub}
R. J. Glauber, in Lectures in Theoretical Physics, Vol. 1, edited by
W. E. Brittin et al. (Interscience, New York, 1959), p.~315;
J. M. Eisenberg and D. S. Kulton, Theory of Meson Interaction with Nuclei
(John Wiley $\&$ Sons, New York, 1980), p.~158.

\bibitem{Lacombe}
M. Lacombe et al., Phys. Lett. B  {\bf 101} (1981) 139.

\bibitem{Jager}
C. W. De Jager, H. De Vries, and C. De Vries, At. Data Nucl. Data
Tables {\bf 14} (1974) 479.


\end{thebibliography}
\end{document}